\documentclass[showpacs,superscriptaddress,aps ]{revtex4}
\usepackage{amssymb}
\usepackage[centertags]{amsmath}
\usepackage{txfonts}
\usepackage{epsfig}
\usepackage{bm}
\usepackage{color}
\usepackage{graphicx,graphics}
\usepackage{multirow}

\begin{document}

\title{$K^{*0}$ and $\Sigma^*$ production in Au+Au collisions at $\sqrt{s_{NN}}=$ 200 GeV and 62.4 GeV}

\author{Kai Zhang}
\affiliation{Department of Physics, Qufu Normal University, Shandong 273165, People's
Republic of China}

\author{Jun Song}
\affiliation{Department of Physics, Jining University, Shandong 273155,
People's Republic of China}

\author{Feng-lan Shao}
\email{shaofl@mail.sdu.edu.cn} \affiliation{Department of Physics, Qufu Normal
University, Shandong 273165, People's Republic of China}

\begin{abstract}
Applying a quark combination model for the hadronization of Quark Gluon Plasma (QGP) and A Relativistic Transport (ART) model for the subsequent hadronic rescattering process, we investigate the production of $K^{*0}$ and $\Sigma^*$ resonances in central Au+Au collisions at $\sqrt{s_{NN}}=$ 200 GeV and 62.4 GeV. The initial $K^{*0}$ produced via hadronization is higher than the experimental data in the low $p_T$ region and is close to the data at 2-3 GeV/c.
We take into account the hadronic rescattering effects which lead to a strong suppression of $K^{*0}$ with low $p_T$ , and find that the $p_T$ spectrum of $K^{*0}$ can be well described.
According to the suppressed magnitude of $K^{*0}$ yield, the time span of hadronic rescattering stage is estimated to be about 13 fm/c at 200 GeV and 5 fm/c at 62.4 GeV. The $p_T$ spectrum of $\Sigma^*$ directly obtained by quark combination hadronization in central Au+Au collisions at 200 GeV is in well agreement with the experimental data, which shows a weak hadronic rescattering effects. The elliptic flow v2 of $\Sigma^*$ in minimum bias Au+Au collisions at 200 GeV and $p_T$ spectrum of $\Sigma^*$ at lower 62.4 GeV are predicted.
\end{abstract}

\maketitle

\section{Introduction}
The short-lived resonances are efficient tools of probing the properties of the hot and dense medium produced in relativistic heavy ion collisions. At RHIC energies, QGP with extremely high energy density is created in primordial collision stage \cite{stock08RHICrev}, and the system undergoes a long time to expand and cool \cite{Kolb0305084nuth}. The lifetime of the resonance is about a few fm/c, which is less than (or roughly the order of) the lifetime of the system formed in heavy ion collisions. After QGP hadronization, but before the interactions of hadrons cease, the initially produced resonances and stable hadrons will undergo a hadronic rescattering stage.  The resonance might be destructed by rescattering with other hadrons and also be regenerated by the collisions of other hadrons,  and decay daughter particles of resonance are kicked by other hadrons causing the signal loss. The physical properties of resonances, e.g. their masses and widths, might be modified by the surrounding medium \cite{Lutz02npa,Shuryak03Medium}. In addition, the yields and momentum spectra of resonances might be changed.  The experimentally reconstructed resonances are the synthetic results of the hadronization and hadronic rescattering effects.

RHIC and SPS  experiments have provided rich data of production of resonances such as $K^*$, $\Sigma^*$ in relativistic heavy ion collisions \cite{KstarRHIC62_200GeV,StrangeBresonance200Gev,kstarAuAuVpp05y,Kstar158GeV}. The relative yield ratios of resonances to stable hadrons are studied experimentally, which also invokes many theoretical explanations \cite{Bleicher02,vogel06,VogelAndBleicher05}. This progress greatly promote our understanding of QGP hadronization mechanism tested against stable hadrons and the hadronic rescattering dynamics, e.g. the time span of hadronic stage and cross sections of various hadronic interaction channels \cite{Rafelski01,Marker02}.

The production mechanism of resonances at QGP hadronization is hard to identify due to the entanglement of hadronization and rescattering effects. RHIC data, e.g. the phenomena of v2 quark number scaling and high $p/\pi$ ratio etc, show that the production of various stable hadrons at QGP hadronization is realized by the combination of constituent (anti-)quarks \cite{Fries:2003prc,Greco2003prc,Hwa:2004prc}. Recently, STAR experiments observed that the v2 of $K^{*0}$, the same as stable particles, follows the constituent quark number scaling rule in the intermediate $p_T$ region \cite{KstarRHIC62_200GeV}. This provides an evidence for the quark (re-)combination/coalescence mechanism of resonance production at hadronization in relativistic heavy ion collisions at RHIC. The K* meson and $\Sigma^*$ baryon are of particular interests due to their very short lifetime (~4fm/c) and strange valence quark content. The experimental data of their midrapidity yields and $p_T$ spectra are all available recently \cite{KstarRHIC62_200GeV,StrangeBresonance200Gev}. A systematical study of this pair of meson and baryon resonance should be able to further test the quark combination mechanism of the hadron production in relativistic heavy ion collisions.

In this paper, we apply a quark (re-)combination model for the hadronization of hot and dense quark matter and ART model \cite{LbaoART95,LbaoART01} for the hadronic rescattering process to study the $K^{*0}$ and $\Sigma^*$  production in central Au+Au  collisions at $\sqrt{s_{NN}}=$200 GeV and 62.4 GeV. The investigation strategy is divided into two steps. Firstly, we compare directly the hadronization results with experimental data to investigate the proportion/magnitude of hadronization exhibited in the final observation.
 Here, as a tool, we use the quark combination model developed by Shandong group (SDQCM) \cite{QBXie1988PRD,Shao2005prc} to deal with QGP hadronization.
  Secondly, we take into account effects of hadronic rescattering and study the entanglement of hadronization and hadronic rescattering effects at RHIC 200 GeV and 62.4 GeV and time span of hadronic stage for the system produced at high RHIC energies.

\section{Initial $K^{*0}$ production via hadronization }
The performance of quark (re-)combination mechanism on explaining the production of various stable hadrons in the intermediate $p_T$ region in relativistic heavy ion collisions is pretty well \cite{Fries:2003prc,Greco2003prc,Hwa:2004prc,ClwKoCM_phi_Omega}. The mechanism can also well describe $p_T$-integrated yields and rapidity spectra of hadrons at RHIC and high SPS energies \cite{Shao2005prc,shao2007prc,CEShao2009PRC,JSong2009MPA}. There are several popular recombination models at RHIC. Quark recombination model \cite{Fries:2003prc,Hwa:2004prc} and parton coalescence model \cite{Greco2003prc} inclusively describe the combination of quarks into hadrons. ALCOR \cite{alcor95} and SDQCM apply the exclusive description. The spirit of quark combination has been extended to various transport, variation and statistic methods of hadron production in relativistic heavy ion collisions \cite{RavaTS07,MHZpfDyCoal07,Alaladyqcm08,Cassing09,Abir09}.

In this paper, we use SDQCM to treat the initial production of various hadrons. Of all the ``on market" combination models, SDQCM is unique for its combination rule which guarantees that mesons and baryons exhaust the probability of all the fates of the (anti)quarks in deconfined color-neutral system at hadronization. The main idea of the combination rule is to line up the (anti)quarks in a one-dimensional order in phase space, e.g., in rapidity, and then let them combine into initial hadrons one by one according to this order \cite{Shao2005prc}. Three (anti)quarks or a quark-antiquark pair in the neighborhood form a (anti)baryon or a meson, respectively. The exclusive nature of the model make it  convenient to predict the $K^{*0}$ and $\Sigma^*$ production on the basis of the reproduction of the yields and momentum spectra of various stable hadrons.

In the meson formation, the relative formation probability of the lowest lying vector meson (V) to pseudo-scalar meson (P) with the same valance quark composition is tuned by the parameter V/P ratio which can not be given from first principles. Spin counting arguments would suggest a 3:1 mixture between vector and pseudoscalar mesons. In Lund string fragmentation, the V/P ratio is taken to be 1 for light mesons and 1.5 for strange mesons, which is based on wave function overlap arguments \cite{And82a}. The resulting  $K^*/K$ yield ratio is about 0.56 which is calculated using PYTHIA 6.4 with default settings \cite{pythia6.4}. The data of $K^{*0}/K^-$ yield ratio in $pp$ collisions at RHIC energies is about 0.35 \cite{kstarAuAuVpp05y}. This will lead to a rough estimation of the V/P ratio by the relationship V/(V+P) =0.35 after taking into account $K^*$ decay, and the resulting V/P ratio is about 0.5.  The SDQCM fit of the value of $K^{*0}/K^-$ in $pp$ reactions gives V/P ratio of 0.45.  The choice of V/P ratio influence directly the predicted abundance of $K^{*0}$ at hadronization, and thus influence the identification of the magnitude of rescattering effects in hadronic stage in explaining the experimental data. Here, we give the prediction of $K^{*0}$ at different V/P values to test its influence quantitatively.

\begin{figure}
  \includegraphics[width=\linewidth]{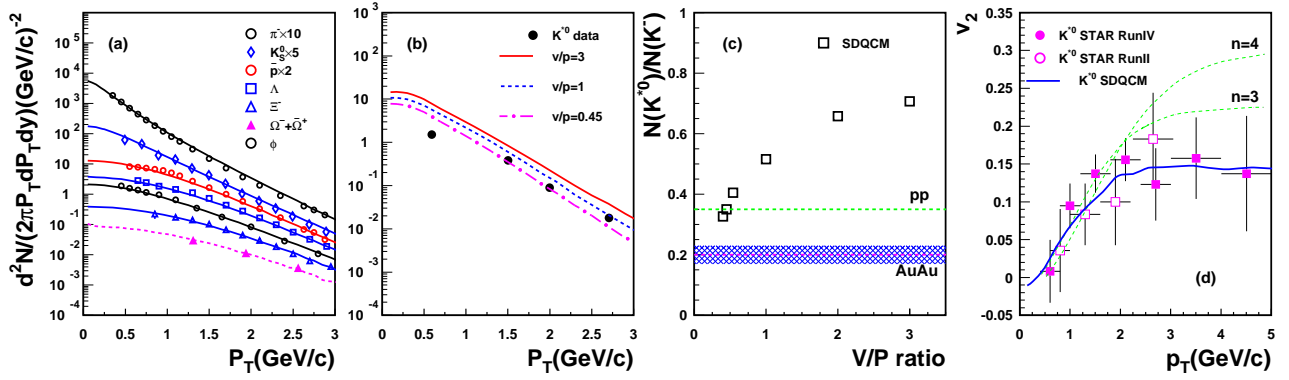}\\
  \caption{Panel (a): The $p_T$ spectra of stable hadrons at midrapidity in central Au+Au collisions at $\sqrt{s_{NN}}=$ 200 GeV. Panel (b): The $p_T$ spectrum of $K^{*0}$. Symbols are the experimental data \cite{abelev:152301,Adams07hyperon,Abelev07phiv2,KstarRHIC62_200GeV} and lines are the results of SDQCM.  Panel (c): Yield ratio of $K^{*0}/K^-$ at midrapidity. Open squares are results by hadronization for different V/P ratios. Experimental data of $pp$ collisions and central Au+Au collisions are shown as dashed line and band area \cite{KstarRHIC62_200GeV}, respectively.  Panel (d): Elliptic flow v2 of $K^{*0}$ as the function of $p_T$ in minimum bias Au+Au collisions at $\sqrt{s_{NN}}=$ 200 GeV. Symbols are the experimental data \cite{KstarRHIC62_200GeV} and the solid line is the result of SDQCM.  Dashed lines marked by $n=3$ and 4 are the guidance of the consequence of aggravating regeneration effects in hadronic stage, which is from Ref. \cite{DongX04v2decay}.}\label{ktar_pt_200GeV}
\end{figure}

The left panel (a) in Fig.\ref{ktar_pt_200GeV} is the $p_T$ spectra of various stable hadrons at midrapidity in central Au+Au at $\sqrt{s_{NN}}=$  200 GeV. Symbols are the experimental data and lines are results of SDQCM in Ref. \cite{CEShao2009PRC} in which the $p_T$ spectra of their anti-particles are also studied in detail.  V/P ratio is taken to be 3 in the calculation. The input of model is the $p_T$ spectra of light and strange quarks just before hadronization, i.e. $f_q(p_T)$ and $f_s(p_T)$,  which are fixed by the data of $\pi^-$ and $K^-$.  Clearly, the transverse momentum spectra of these stable hadrons can be self-consistently explained by two quark $p_T$ spectra via combination.  This means that constituent quark degrees of freedom play a dominated role in the production of these thermal hadrons.

The middle panel (b) in Fig.\ref{ktar_pt_200GeV} shows midrapidity $p_T$ spectrum of $K^{*0}$ produced by hadronization based on the left panel results and parameters (quark spectra and V/P ratio). The V/P ratio is sensitive to the  $K^{*0}$ yield and is relatively less sensitive to the slope of $K^{*0}$ $p_T$ spectrum. The $K^{*0}$ spectrum at other V/P values, e.g. 1 and 0.45, are all presented.  Here, we stress that even at V/P=1 and 0.45 the model can still reproduce the $p_T$ and rapidity spectra of stable hadrons shown left.  It is because the V/P ratio does not alter the nature of hadron formation, and it just changes the decay contributions of resonance to stable hadrons. We find that the slope of $K^{*0}$ $p_T$ spectrum  in 1.5-3.0 GeV/c is roughly consistent with the data but in the low $p_T$ region the directly produced $K^{*0}$ by combination is above the data even for V/P=0.45.

The right panel (c) in Fig.\ref{ktar_pt_200GeV} is the yield ratio $N(K^{*0})/N(K^-)$ at midrapidity for different V/P values. Because kaon and $K^*$ both contain same valence quarks, their ratio roughly cancel the effect of strangeness enhancement in relativistic heavy ion collisions and thus is sensitive to the mechanism of hadron production. The data of $N(K^{*0})/N(K^-)$ is $0.2\pm0.03\pm0.03$ in central Au+Au collisions and is $0.34\pm0.01\pm0.05$ in minimum bias $pp$ reactions at  $\sqrt{s_{NN}}=$ 200 GeV  \cite{KstarRHIC62_200GeV}, respectively. We find that the calculated ratio $N(K^{*0})/N(K^-)$ at V/P =3 and 1 are higher than the data of Au+Au collisions which is shown as the band area. The result of V/P=0.45  is consistent with the data of $pp$ reactions shown as dashed line but is also higher than the data of Au+Au collisions. The over-prediction of hadronization in the low $p_T$ region indicates the necessity of  hadronic rescattering for the suppression of $K^{*0}$ yield.

To further manifest the hadronization effect in final state observation, we present in panel (d) in Fig.\ref{ktar_pt_200GeV} the elliptic flow v2 of $K^{*0}$ at midrapidity in minimum bias Au+Au collisions at $\sqrt{s_{NN}}=$ 200 GeV. This result is calculated using the extracted quark v2 in the previous study of v2 of various stable hadrons in Ref. \cite{qcmv2}. V/P ratio does not influence the predicted v2 which is irrespective of particle abundance. One can see that the calculated $K^{*0}$ v2, shown as solid line, is in well agreement with the experimental data. The dashed lines are the parameterization of hadronic v2 as the function of number of constituent quarks $n$ in Ref. \cite{DongX04v2decay}. $n=3$ and 4 are the guidance of the consequence of aggravating regeneration effects in hadronic stage.  The agreement of our result with the data means that the $K^{*0}$ mesons observed experimentally mainly come from the hadronization.

\section{Hadronic rescattering effects on $K^{*0}$ production }
The subsequent hadronic rescattering stage after hadronization is simulated by A Relativistic Transport (ART) model \cite{LbaoART95,LbaoART01} which includes baryon-baryon, baryon-meson, and meson-meson elastic and inelastic scatterings. We apply the code in AMPT event generator V2.25t3 \cite{LinAMPT05} which has provided proper extension to higher RHIC energies from original SPS energies. The time span of hadronic stage is important for the hadronic rescattering effects such as the suppression magnitude of $K^{*0}$ yield. In transport theory, the rescattering time should be infinite in principle. In practice, one would like to choose a finite rescattering time, and this choice, to some extent, is  similar to the decouple criteria of kinetic freeze-out in hydrodynamic theory of heavy ion collisions. In this work, we treat the rescattering time as an adjustable parameter and study how long the time span of hadronic phase is favored by the experimental data when the initial produced $K^{*0}$ is fixed.

\begin{figure}
  \includegraphics[width=6cm]{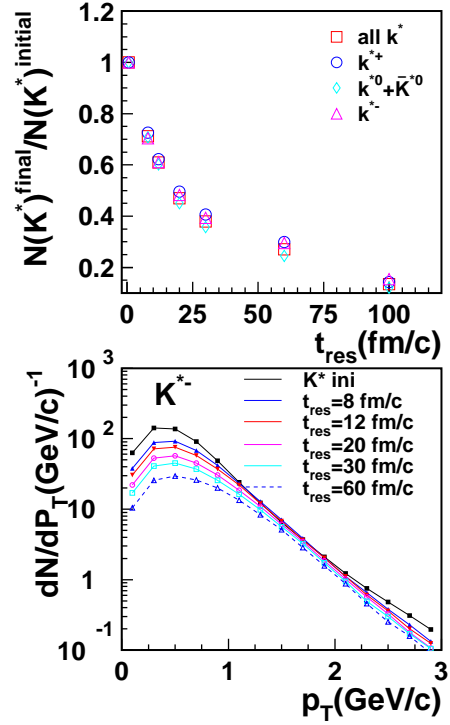}\\
  \caption{Top panel: The survived ratio of $K^*$yield as the function of hadronic rescattering time $t_{res}$ in central Au+Au collisions at $\sqrt{s_{NN}}=$ 200 GeV. Bottom panel: The survived $K^{*0}$ as the function of $p_T$ at different $t_{res}$. }\label{survive_200_ampt}
\end{figure}

\begin{figure}
  \includegraphics[ width=6cm]{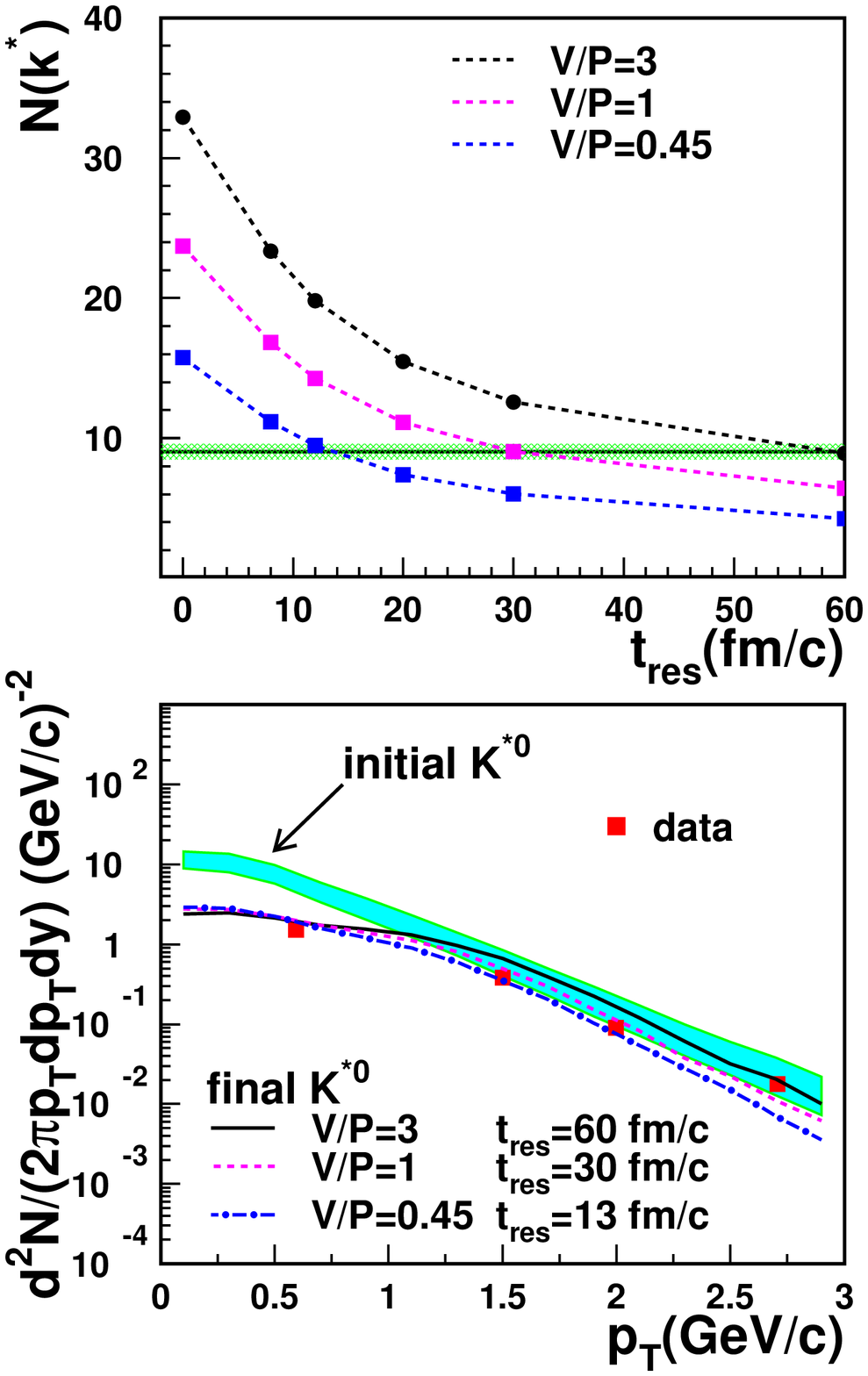}\\
  \caption{Top panel: The final state $K^*$yield at midrapidity as the function of hadronic rescattering time $t_{res}$ in central Au+Au collisions at $\sqrt{s_{NN}}=$ 200 GeV. The symbols connected with lines are calculation results corresponding three differen V/P ratios at hadronization. The experimental data are shown as the band area \cite{KstarRHIC62_200GeV}. Bottom panel: The $p_T$ spectra of final state $K^*$ for three V/P values with the corresponding rescattering times for yield reproduction in central Au+Au collisions at 200 GeV. The experimental data are shown as symbols \cite{KstarRHIC62_200GeV}. The band area shows the result of $K^*$ just after hadronization. }\label{final_kstar_200gev}
\end{figure}

Fig.\ref{survive_200_ampt} top panel shows the survived ratio of $K^*$ abundance $N(K^*)^{final}/N(K^*)^{initial}$ after hadronic rescattering stage as the function of rescattering time $t_{res}$. Here $N(K^*)^{initial}$ is the number of  $K^*$ just after hadronization and  $N(K^*)^{final}$ is the number of  $K^*$ that can be experimentally reconstructed after hadronic rescatterings.  $K^*$ mesons are reconstructed from their hadronic decay channels using pion-kaon invariant mass analysis, and the survived $K^*$ incorporates all effects of destruction, signal loss and regeneration. One can see that the number of survived $K^*$ almost exponentially decreases with the rescattering time. The time of $K^*$ number reducing by half is about 20 fm/c. Fig.\ref{survive_200_ampt} bottom panel  shows the number of survived $K^{*0}$ after hadronic rescattering stage as the function of transverse momentum $p_T$ at different rescattering time $t_{res}$. We find that the suppression of $K^{*0}$ caused by hadronic rescattering effects is strong in low $p_T$ region. This is qualitatively consistent with the result of UrQMD transport model \cite{VogelAndBleicher05,Kstar158GeV}.

Taking into account of effects of hadronic rescatterings on the yield suppression and spectra distortion, we obtain the yield and $p_T$ spectrum of final state $K^{*0}$ comparable to experimental data. The top panel in Fig.\ref{final_kstar_200gev} shows the yield of final survived $K^{*0}$ at midrapidity as the function of rescattering time $t_{res}$ in central Au+Au collisions at $\sqrt{s_{NN}}=$ 200 GeV. The directly produced $K^{*0}$ for different V/P ratios at hadronization is taken to be the starting point of the hadronic rescattering stage. For V/P=3 the rescattering time needed to reproduce the experimental data is about 60 fm/c while for V/P=1 the needed time is about 30 fm/c. For V/P=0.45 favored by $pp$ $N(K^{*0})/N(K^-)$ data the rescattering time is about 13 fm/c.
 This value is consistent with the typical lifespan of $13\pm3$ fm/c in UrQMD simulation of hadronic rescattering stage \cite{Bleicher02}. The estimation of time span between chemical and thermal freeze-out by a thermal model using an additional pure rescattering phase is $2.5^{+1.5}_{-1}$ fm/c from the analysis of $K^{*0}/K^-$  yield ratio \cite{Rafelski01,VogelAndBleicher06proc}.
The bottom panel in Fig.\ref{final_kstar_200gev} shows the $p_T$ spectrum of final state $K^{*0}$ at midrapidity at three V/P values with the corresponding rescattering times for yield reproduction in central Au+Au collisions at 200 GeV.   We find that the strong suppression of low $p_T$ $K^{*0}$ in hadronic rescattering process offset the over-predication of initial $K^{*0}$ by hadronization shown as the band area, and this leads to an obviously improved description of $p_T$ spectrum of $K^{*0}$.

\begin{figure}
  \includegraphics[width=16cm]{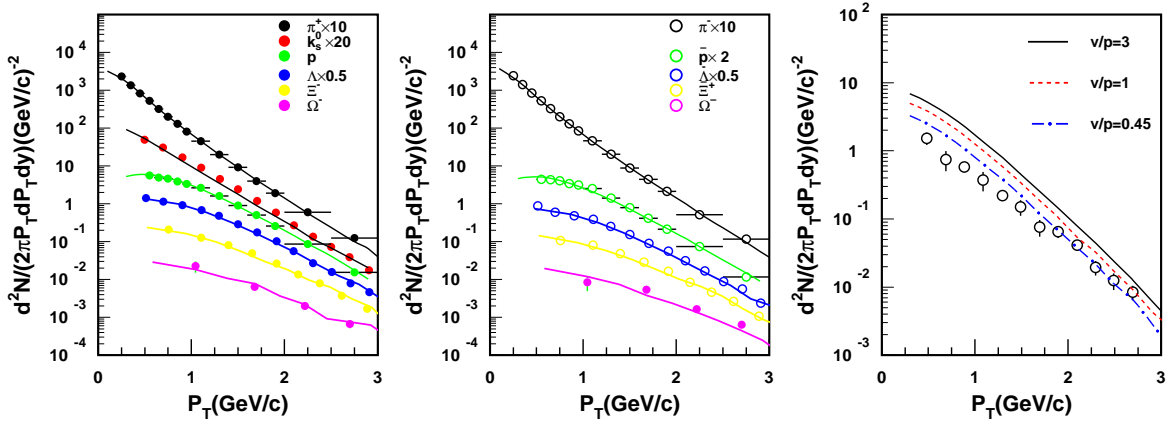}\\
  \caption{The $p_T$ spectra of stable hadrons (left panel), their anti-hadrons (middle panel) and $K^{*0}$ resonance (right panel) at midrapidity in central Au+Au collisions at $\sqrt{s_{NN}}=$ 62.4 GeV. Symbols are the experimental data \cite{piproton62gev,multH62.4GeV} and lines are results of SDQCM. }\label{hpt_62gev}
\end{figure}

The $K^{*0}$ data at different collision energies enable the further test of hadronization mechanism and the investigation of the energy dependence of hadronic rescattering effects. In Fig.\ref{hpt_62gev}, we present the $p_T$ spectra of various stable hadrons and $K^{*0}$ resonance at midrapidity in central Au+Au collisions at $\sqrt{s_{NN}}=$ 62.4 GeV. Symbols are the experimental data and lines are the results of SDQCM. The $p_T$ spectra of constituent quarks at hadronization are taken to be the thermal exponential pattern $\exp(-m_T /T_s)$. The slope parameter $T_s$ is taken to be 0.31 GeV for strange quarks and 0.29 GeV for light quarks, which are smaller than those at 200 GeV \cite{CEShao2009PRC}. The numbers and rapidity spectra of constituent quarks and antiquarks are fixed by the experimental data of pion and kaon rapidity spectra \cite{Arsene62gev_y_spectra}. One can see that, similar to Au+Au 200 GeV, the model results of various stable hadrons are in well agreement with the data. Then we can predict the $p_T$ spectrum of initial $K^{*0}$ just after hadronization, and the results are presented in the right panel in Fig.\ref{hpt_62gev}. The degree of the agreement between the model result and experimental data is also similar to that in Au+Au 200 GeV. The calculated $K^{*0}$ yield densities in low $p_T$ region exceed the data and become to close to the data as $p_T$ rises to 2-3 GeV/c. This indicates the influence of the hadronic rescattering on $K^{*0}$ production is still significant at intermediate RHIC energy.

\begin{figure}
  \includegraphics[width=6cm]{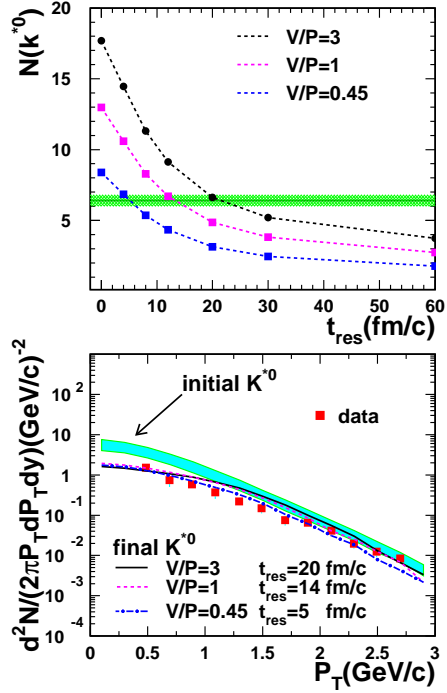}\\
  \caption{Top panel: The final state $K^*$ yield at midrapidity as the function of hadronic rescattering time $t_{res}$ in central Au+Au collisions at $\sqrt{s_{NN}}=$ 62.4 GeV. The symbols connected with lines are calculation results corresponding three differen V/P ratios at hadronization. The experimental data are shown as the band area \cite{KstarRHIC62_200GeV}. Bottom panel: The $p_T$ spectra of final state $K^*$ for three V/P values with the corresponding rescattering times for yield reproduction in central Au+Au collisions at 62.4 GeV. The experimental data are shown as symbols \cite{KstarRHIC62_200GeV}. The band area shows the result of $K^*$ just after hadronization. }\label{final_kstar_62gev}
\end{figure}

The top panel in Fig.\ref{final_kstar_62gev} shows the yield of final survived $K^{*0}$ at midrapidity as the function of rescattering time $t_{res}$ in central Au+Au collisions at $\sqrt{s_{NN}}=$ 62.4 GeV.  For V/P=3 the rescattering time needed to reproduce the experimental data is about 20 fm/c while for V/P=1 the needed time is about 14 fm/c and for V/P=0.45 favored by $pp$ N($K^*$)/N($K^-$ ) data the time is about 5 fm/c. It is found that the needed time at 62.4 GeV is about half of that at 200 GeV.

The bottom panel in Fig.\ref{final_kstar_62gev} presents the $p_T$ spectrum of final state $K^{*0}$ at midrapidity at three V/P values with the corresponding rescattering times for yield reproduction in central Au+Au collisions at 62.4 GeV. The strong suppression in low $p_T$ region in hadronic rescattering process offset the over-predication of initial $K^*$by hadronization shown as the band area, and this leads to an improved  description of $p_T$ spectrum of $K^{*0}$.

\section{$\Sigma^*$ production by hadronization at RHIC}
\begin{figure}
  \includegraphics[width=16cm]{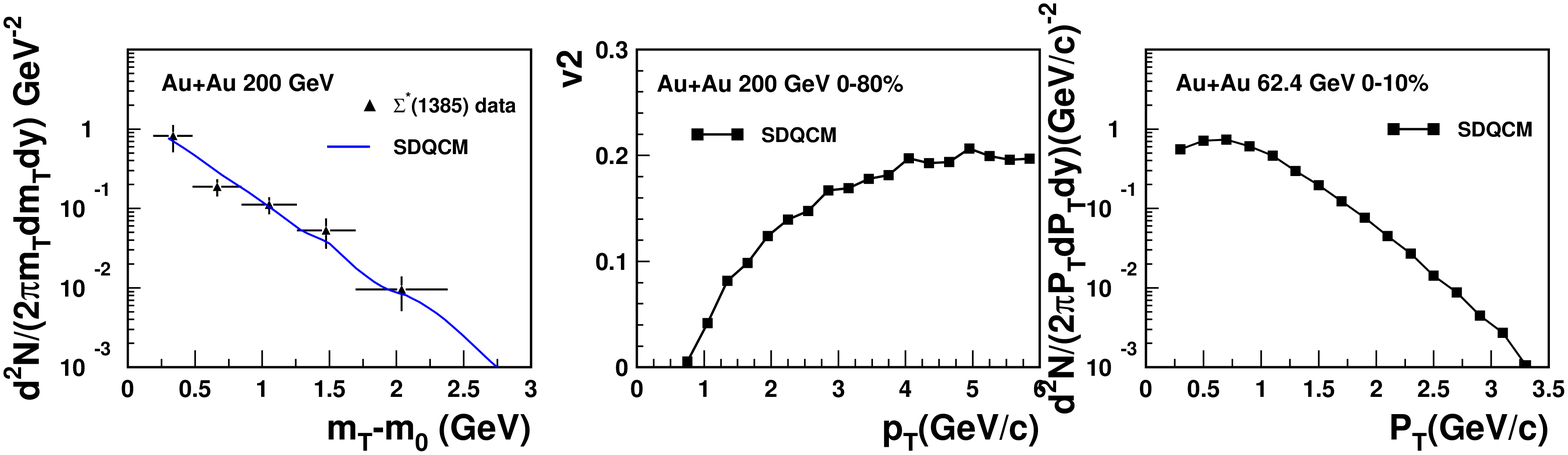}\\
  \caption{Left panel: The $p_T$ spectrum of $\Sigma^*$ in central Au+Au collisions at $\sqrt{s_{NN}}=$ 200 GeV. Symbols are the experimental data \cite{StrangeBresonance200Gev} and the line is the result of SDQCM. Middle panel: The prediction of elliptic flow v2 of $\Sigma^*$ in minimum bias Au+Au collisions at 200 GeV. Right panel: The prediction of $p_T$ spectrum of $\Sigma^*$  in central Au+Au collisions at $\sqrt{s_{NN}}=$ 62.4 GeV. }\label{sigam_pt_v2}
\end{figure}

The $\Sigma^*$ hyperon has a very short lifetime (4.5 fm/c). The measurement of STAR Collaboration found that the $\Sigma^*/\Lambda$ yield ratio at midrapidity in central Au+Au collisions is nearly the same as that in $pp$ collision at $\sqrt{s_{NN}}=$ 200 GeV \cite{StrangeBresonance200Gev}. This indicates that the net effects of rescattering loss and rescattering gain  in hadronic stage are very small.  This poses serious constraint on cross sections of various reaction channels of collision gain and collision loss. The small net effect of hadronic rescattering has two possibilities. The first is that both the effect of collision gain and that of collision loss on $\Sigma^*$ yield are large but they offset with each other. The $p_T$ spectrum of  $\Sigma^*$ might change during hadronic rescattering in this case. The second is that two components are all small and $p_T$ spectrum of $\Sigma^*$ should not be changed dramatically compared with the result of hadronization.  Here, using the quark spectra fixed by stable hadrons in Fig.\ref{ktar_pt_200GeV},
 we use SDQCM to study the yield and $p_T$ spectrum of  $\Sigma^*$ at midrapidity  in central Au+Au collisions at $\sqrt{s_{NN}}=$ 200 GeV.  The yield density $dN/dy$ of  $\Sigma^* + \overline{\Sigma}^*$ at midrapidity just after hadronization is 9.0, and the data of STAR Collaboration is $9.3\pm1.4\pm1.2$ \cite{StrangeBresonance200Gev}. Here $\Sigma^{*}$ represents  $\Sigma^{*+}+\Sigma^{*-}$.
 The result of $p_T$ spectrum of   $\Sigma^{*}$ just after hadronization is presented in left panel in Fig.\ref{sigam_pt_v2} and is compared with the experimental data. We find that the agreement between hadronization results and the data is well. The result favors the second case. We further predict the elliptic flow v2 of $\Sigma^*$ in minimum bias Au+Au collisions at 200 GeV and the result is shown in middle panel in Fig. \ref{sigam_pt_v2}.
 It is well known that the elliptic flow is sensitive to the mechanism of hadron production.   The elliptic flow of hadrons formed via quark combination mechanism follows the constituent quark number scaling rule. The phenomenological regeneration of $\Sigma^*$ by $\Lambda+\pi \rightarrow \Sigma^*$ will result in higher v2.
To test the effects of hadronic rescattering on $\Sigma^*$ production at RHIC energies, we further predict the $\Sigma^*$ $p_T$ spectrum in central Au+Au collisions at 62.4 GeV.  The yield density of $\Sigma^*$ at midrapidity is 4.3 just after hadronization.

\section{summary}
In this paper, we have used SDQCM model for the QGP hadronization and ART model for the subsequent hadronic rescattering process to investigate the production of $K^{*0}$  and $\Sigma^*$ resonances in central Au+Au collisions at $\sqrt{s_{NN}}=$ 200 GeV and 62.4 GeV.
The initial $K^{*0}$ produced by quark combination mechanism is higher than the experimental data in the low $p_T$ region and is close to data at 2-3 GeV/c.
In the subsequent hadronic rescattering stage, the number of $K^{*0}$  that can be reconstructed experimentally exponentially decreases with the increasing rescattering time, and the suppression of $K^{*0}$ yield focuses on low $p_T$ region, which offsets the over-prediction of hadronization. Therefore the quark combination hadronization plus hadronic rescattering  can provide a well description of the experimental data of $K^{*0}$ production.
According to the suppression magnitude of $K^{*0}$ yield, the time span of hadronic rescattering stage is estimated.  V/P ratio at hadronization is important for the extraction of hadronic rescattering time from the data of $K^{*0}$ yield.   For V/P =0.45  based on the data of $K^{*0}/K^-$ in $pp$ reaction, the time span of hadronic rescattering stage is about  13 fm/c at 200 GeV and 5 fm/c at 62.4 GeV.   Higher V/P ratio leads to the  longer rescattering time.
The yield density and $p_T$ spectrum of $\Sigma^*$ directly from quark combination hadronization in central Au+Au collisions at 200 GeV is found to be in well agreement with the experimental data, which indicates a weak hadronic rescattering effects. To make a further test, we predict the v2 of $\Sigma^*$ in minimum bias Au+Au collisions at 200 GeV and $p_T$ spectrum of $\Sigma^*$ at lower 62.4 GeV for the comparison with future STAR data .

\section*{ACKNOWLEDGMENTS}
The authors thank X. B. Xie, Z. T. Liang and G. Li for helpful discussions.  The work is supported in part by the National Natural Science Foundation of China under grant 11175104 and 10947007, and by  the Natural Science Foundation of Shandong Province, China under grant ZR2011AM006.

\end{document}